\documentclass[envcountsame,oribibl,a4paper,UKenglish]{llncs}
\pdfoutput=1
\usepackage{
amsmath
}
\usepackage{amssymb}
\usepackage{graphicx}

\newcommand{\defemph}[1]{\textbf{\emph{#1}}}

\title{On Searching for Small Kochen-Specker\\ Vector Systems
(extended version)%
\thanks{This work is based on the first author's Master's
  thesis~\cite{Arends09alowerbound}. A shorter version of this paper
  previously appeared as~\cite{AOW11a}.}}

\author{Felix Arends\inst{1} \and Jo\"el Ouaknine\inst{2} 
\and Charles W. Wampler\inst{3}}

\institute{Google Germany GmbH\\
          \email{felix.arends@gmx.de}
\and Department of Computer Science, Oxford University, UK\\
          \email{joel@cs.ox.ac.uk}
\and Department of Mathematics, University of Notre Dame, USA\\
          \email{charles.w.wampler@gm.com}}

\begin{document}

\maketitle

\begin{abstract}
Kochen-Specker (KS) vector systems are sets of vectors in
$\mathbb{R}^3$ with the property that it is impossible to assign 0s
and 1s to the vectors in such a way that no two orthogonal vectors are
assigned 0 and no three mutually orthogonal vectors are assigned
1. The existence of such sets forms the basis of the Kochen-Specker
and Free Will theorems. Currently, the smallest known KS vector system
contains 31 vectors. In this paper, we establish a lower bound of 18
on the size of any KS vector system. This requires us to consider a
mix of graph-theoretic and topological embedding problems, which we
investigate both from theoretical and practical angles. We propose
several algorithms to tackle these problems and report on extensive
experiments. At the time of writing, a large gap remains between the
best lower and upper bounds for the minimum size of KS vector
systems.\\

\end{abstract}

\section{Introduction}

In a recent, thought-provoking paper, John H. Conway and Simon
Kochen demonstrate that ``\emph{if [\ldots] there exist any experimenters
with a modicum of free will, then elementary particles must have their
own share of this valuable commodity}''~\cite{conway-2006}.
More precisely, Conway and Kochen consider so-called `spin-1'
particles (such as photons) whose `spin' (a physical property) can be
measured along any given direction. The squared outcome of such
measurements is always either 0 or 1. Conway and Kochen's Free Will
theorem asserts that, if an experimenter can choose the direction
along which to perform a spin-1 experiment freely (i.e., in a way that
is not determined by the past), then the response of the spin-1
particle to such an experiment is also not determined by the past.

This theorem rests on three basic axioms of quantum mechanics and
relativity, the most crucial of which (for our purposes) is the
following:\vspace{1ex}

\noindent
\textbf{The SPIN Axiom~\cite{conway-2006}.}
\emph{Measurements of the squared components of spin of a spin-1 particle in
three orthogonal directions always yield the outcomes 1, 0, 1 in some
order.}\footnote{As pointed out in \cite{conway-2006}, such
measurements `commute', so the order in which they are performed does
not matter.}

The SPIN axiom not only follows from the postulates of quantum
mechanics, but has also been verified
experimentally~\cite{PhysRevLett.90.250401}. This axiom alone already
gives rise to what is known as the `Kochen-Specker
paradox'~\cite{conway-2008}: if the response of a spin-1 particle to
any conceivable spin measurement were predetermined prior to the actual
measurement, then those responses would define a function from the
unit sphere in three dimensions to the set $\{0, 1\}$, satisfying the
so-called 101-property: any three points on the sphere with mutually
orthogonal position vectors must be assigned the values 1, 0, 1 in
some order. The Kochen-Specker paradox---which is in fact a
mathematical theorem---is that no such function exists.

The impossibility of such `101-functions' can be proved by exhibiting
a finite set of points on the sphere on which such functions cannot be
defined. The first such set, discovered by Kochen and Specker more
than forty years ago, contained 117
points~\cite{Kochen67theproblem}. Subsequent sets, usually referred to
as `records'~\cite{pavicic-2005-38}, cut this number down to 33 and
then 31~\cite{Peres}. The latter is the size of the smallest known
`Kochen-Specker vector system', discovered approximately 20 years ago
by Conway and Kochen.

As pointed out in~\cite{pavicic-2005-38}, finding small Kochen-Specker
vector systems has both theoretical and practical motivations. Conway
himself has stressed the problem on several occasions whilst giving
public lectures on the Free Will theorem. The work we describe here
reports on some partial progress in this endeavour; our main result is
that any Kochen-Specker vector systems must contain at least 18
vectors (Thm.~\ref{lowerbound18}). Achieving this bound required us to
consider a mix of graph-theoretic and topological embedding problems,
for which we devised and analysed a number of algorithms. In addition,
we establish bounds on the theoretical complexity of some of the
principal problems involved (Thms.~\ref{101thm} and
\ref{embedding_complexity}), and also show that the key task of
checking canonicity in the 30-year-old Colbourn-Read orderly graph
enumeration algorithm~\cite{colbourn-1979a} cannot belong to NP---and
much less to P---unless NP = co-NP (Thm.~\ref{hardnesscanonicity}).

Unfortunately, it would appear that narrowing the gap between the best
lower and upper bounds for the minimum size of KS vector systems
remains a formidable challenge, and significant progress in this area
will likely require substantially new ideas.

The work most closely related to ours is that of Pavi\v{c}i\'{c}
\emph{et al.}~\cite{pavicic,pavicic-2005-38}. They consider
higher-dimensional generalisations of the problem treated in this
paper, but their formulation and results are incomparable to
ours. We return to the differences between our approach and theirs in
Sec.~\ref{secKS}; we also refer the reader to
\cite{pavicic,Arends09alowerbound} for a more thorough discussion of
the matter.\vspace{1ex}

\noindent
\textbf{Different Standards of Proof.} Due to the mixed
discrete/continuous aspects of the problems considered in this paper,
it is important to pay special attention to the nature of the proofs
involved. The usual kind of proof is \emph{mathematical}. Along with
such proofs, we also present several results that have
\emph{computer-aided} proofs, in which extensive calculations were
carried out by computer. A third category of results could be deemed
to have \emph{numerical} proofs, by which we mean that a computer
program was used and floating-point arithmetic was involved in a way
that cannot be guaranteed to be entirely accurate. It is reasonable to
assume that the results thus obtained are very likely to be correct,
but it remains conceivable that a highly ill-conditioned initial
problem could lead to an incorrect answer.

It should be stressed that in this paper, we do not consider numerical
calculations and proofs to be sufficiently reliable to be fully
adequate; however, we have frequently made use of numerical techniques
as heuristics in order to guide and accelerate the search for
\emph{computer-aided} proofs of our main results. The latter
should be viewed as having the same strength as traditional
pencil-and-paper, mathematical proofs.

\section{Kochen-Specker Vector Systems}
\label{secKS}

Kochen-Specker vector systems can be represented in multiple ways. In
the Introduction, we have implicitly described such systems as certain
finite sets of points on the surface of the sphere $\mathbb{S}^2$. In
fact, an immediate consequence of the SPIN axiom is that squared-spin
measurements along opposite directions necessarily yield the same
outcome, so that it is sensible to identify antipodal
points. Accordingly, let us therefore define a \defemph{vector system}
as a finite subset of the open northern hemisphere $\mathbb{H}^2 =
\{(x,y,z) : x^2+y^2+z^2 =1 \mathrm{\ and\ }
z>0\}$.\footnote{Dispensing entirely with the equator simplifies
somewhat our technical development later on; it is harmless since any
finite set of points on the sphere can always be rigidly rotated so as
to avoid the equator.}  Alternatively, vector systems can be
represented as finite subsets of the projective plane $\mathbb{P}^2$,
and also as finite sets of points on the surface of a cube, where once
again we identify antipodal points.  We variously make use of all
three of these representations in the rest of this paper.

A vector system $\mathcal{K} \subseteq \mathbb{H}^2$ is said to be
\defemph{101-colourable} if it is possible to assign either 0 or 1 to
each vector in $\mathcal{K}$ such that (i)~no two orthogonal vectors
are both assigned 0, and (ii)~no three mutually orthogonal vectors are
all assigned 1.

Finally, a \defemph{Kochen-Specker (KS) vector system} is a vector
system that is not 101-colourable. The \emph{size} of such a system is
simply the number of vectors it contains.

\begin{figure}
  \begin{center}
    \includegraphics{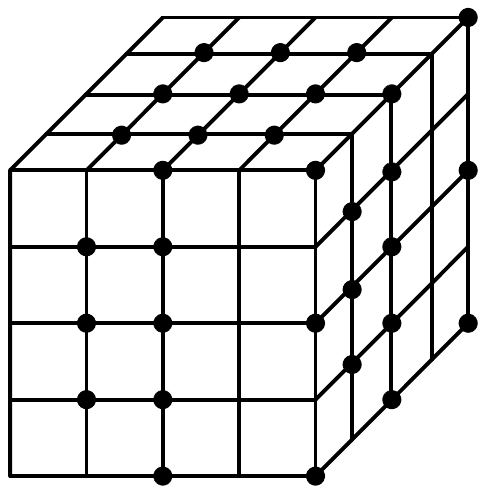}
  \end{center}
  \caption{\label{fig_conway2} A visual representation of Conway and
      Kochen's KS vector system of size 31.}
\end{figure}

At the time of writing, the smallest known KS vector system is still
the one discovered approximately twenty years ago by Conway and
Kochen~\cite{Peres}. The 31 vectors of this system can be represented
as lying on a cubic grid centered at the origin, as depicted in
Fig.~\ref{fig_conway2}. Note that orthogonality relationships among
the vectors are easily inferred through elementary geometry, thanks to
the regularity of the grid. Non-101-colourability can be established
by (somewhat tedious) case analysis.

Note that the colourability conditions~(i) and (ii) as given above,
although in appearance stronger than the SPIN axiom, are implicitly
equivalent to it. For instance, while the SPIN axiom, strictly
speaking, asserts nothing about a vector system consisting of exactly
two orthogonal vectors, it implicitly requires one of these vectors to
be assigned 1, since if both were assigned 0 one could derive a
contradiction by considering a third vector orthogonal to the other
two. 

This seemingly innocuous observation has consequences for the way in
which KS vector systems are built and measured.  Pavi\v{c}i\'{c}
\emph{et al.}~\cite{pavicic-2005-38} and Larsson~\cite{LAR02}, for
example, require every pair of orthogonal vectors to belong to a
triple of mutually orthogonal vectors, invoking a strict application
of the SPIN axiom. Following this convention, they argue that the KS
vector system depicted in Fig.~\ref{fig_conway2} should be viewed as
having size \emph{51} rather than 31 (cf.~\cite{pavicic-2005-38}):
indeed, this system as represented above contains 20 pairs of orthogonal
vectors without a third orthogonal vector present.

In contrast, our own conventions---following, among others,
\cite{Kochen67theproblem,Peres91twosimple,Peres,BUB96,conway-2006,conway-2008}---are
predicated on colourability conditions~(i) and (ii) as given
earlier. For further discussion on the matter, we refer the reader
to~\cite{pavicic-2005-38} and \cite{Arends09alowerbound}.\vspace{1ex}

\noindent
\textbf{Vector Systems and Graphs.} Any vector system $\mathcal{K}$ gives
rise to an associated undirected graph $G_{\mathcal{K}}$, the vertices
of which are the vectors of $\mathcal{K}$, with an edge between two
vertices iff the corresponding vectors are orthogonal. In other words,
$G_{\mathcal{K}} = (V,E)$, where $V = \mathcal{K}$ and $E =
\{\{\vec{u},\vec{v}\} : \vec{u},\vec{v} \in \mathcal{K} \mathrm{\ and\
} \vec{u} \cdot \vec{v} = 0\}$.

We define 101-colourability for graphs in the obvious way: assignment
of 0 or 1 to the vertices in such a way that (i)~no two adjacent
vertices are both assigned 0, and (ii)~no triangle (3-clique) is
assigned all 1s. Clearly, $\mathcal{K}$ is a KS vector system iff
$G_{\mathcal{K}}$ is not 101-colourable.

Of course, an arbitrary graph $H$ may not correspond to any realisable
(3-dimensional) vector system: the orthogonality constraints
corresponding to graph edges may fail to be simultaneously
satisfiable. Let us define a graph $H$ to be \defemph{embeddable} if
there exists some vector system that it corresponds to. More
precisely, we ask that there be a vector system $\mathcal{K}$ that can
be put in one-to-one correspondence with the vertices of $H$ in such a
way that adjacent vertices are mapped to orthogonal vectors. Note that
we do \emph{not} require that non-adjacent vertices should go to
non-orthogonal vectors; this relaxation simplifies the
embeddability-checking process, discussed in
Sec.~\ref{chap_embeddability}. However, it is necessary for distinct
vertices to go to distinct vectors. Formally, $H = (V,E)$ is
embeddable if it has a supergraph $H' = (V,E')$ over the same set of
vertices such that $H'$ is isomorphic to $G_{\mathcal{K}}$ for some
vector system $\mathcal{K}$. 

Finding a small KS vector system therefore corresponds to finding a
small graph that is both not 101-colourable and embeddable, and
accordingly this is the approach we have followed and report on in
this paper.

Note that any graph containing a square (4-cycle) is unembeddable:
indeed, orthogonality constraints would force a pair of opposite
vertices of the square to be mapped to collinear (i.e., identical)
vectors. Accordingly, we shall mainly focus on \emph{square-free}
graphs in the remainder of this paper. This turns out to be a fairly
powerful restriction: all square-free graphs with 9 or fewer vertices
are embeddable, while there are only two distinct (up to isomorphism)
unembeddable square-free graphs with 10
vertices~\cite{Arends09alowerbound}.

A second interesting observation about embeddable graphs is the
following:
\begin{proposition}
\label{prop4colourable}
Any embeddable graph is 4-colourable.
\end{proposition}

To see this, consider an embeddable graph $G$ and let $\mathcal{K}$ be
an embedding of it as a vector system in the hemisphere
$\mathbb{H}^2$.  Partition $\mathbb{H}^2$ into four quadrants as
delineated by the $xz$-plane and the $yz$-plane, and colour each
vector of $\mathcal{K}$ according to the quadrant it lies in. Since
vectors belonging to the same quadrant cannot be mutually orthogonal,
corresponding vertices of $G$ cannot be adjacent. Thus the quadrant
colouring of $\mathcal{K}$ gives rise to a valid 4-colouring of $G$.
\qed

As the next result indicates, 101-colourability is in theory an
expensive condition to check, even when restricting to graphs that are
both square-free and 4-colourable. In practice, however, our SAT-based
colourability checker (implemented using MiniSat~2.0 \cite{minisat})
was systematically able to decide 101-colourability of graphs having
at most 30 vertices in microseconds. In fact, experiments with graphs
having hundreds of vertices also completed well within a millisecond.

\begin{theorem}
Deciding whether a square-free 4-col\-our\-able graph is
101-colourable is NP-complete.
\label{101thm}
\end{theorem}

Membership in NP is obvious, whereas hardness can be shown by reduction
from 3-colourability of an arbitrary graph $G$: replace every vertex
of $G$ by a triangle (which can be 101-coloured in precisely three
ways) and replace each edge of $G$ by a gadget ensuring that the
corresponding `adjacent' triangles are coloured differently. This can
be done whilst ensuring that the resulting graph is both square-free
and 4-colourable---details can be found in~\cite{Arends09alowerbound}.
\qed

Finally, note that every 3-colourable graph is automatically
101-col\-our\-able.

\section{Embeddability}
\label{chap_embeddability}

In this section, we examine the problem of determining whether a given
(square-free) graph is embeddable or not. 

It is fairly straightforward to see that embeddability queries can be
phrased in the existential theory of the reals: given a graph $G$,
postulate a triple of real variables $(x,y,z)$ for every vertex of
$G$, and express the various constraints using polynomial equalities
and inequalities. For example, $x^2 + y^2 + z^2 =1$ and $z>0$ together
ensure that the corresponding vector should lie in the hemisphere
$\mathbb{H}^2$. Orthogonality constraints are likewise expressed by
setting the relevant dot products equal to zero, and so
on. Embeddability of the graph $G$ therefore corresponds to
solvability of this constraint system over the reals.

It is plain that the constraints can be constructed in polynomial
time. Since the existential theory of the reals has polynomial space
complexity~\cite{62257,141387}, we have:

\begin{theorem}
\label{embedding_complexity}
Graph embeddability can be decided in PSPACE\@.
\end{theorem}

For a graph with 30 vertices, the corresponding constraint system
requires 90 real variables (or rather, assuming the graph has at least
one triangle, 81 variables since we can quotient out rotational
symmetries by fixing the vectors associated with one of its
triangles). Unfortunately, current real arithmetic solvers cannot in
practice handle systems containing more than just a handful of real
variables. Thm.~\ref{embedding_complexity} is therefore mainly of
theoretical interest at the present time.

Our next observation is that, given a graph $G$, one can in fact
construct a single multivariate polynomial $P_G$ over the reals such
that $G$ is embeddable iff $P_G$ has a root. We simply extend the
above approach by transforming inequalities into equalities, through
the use of auxiliary variables, and conjoining multiple equalities
into a single one via a standard squaring trick. For example, the
inequality $z > 0$ is equivalent to the conjunction of the equalities
$uv = 1$ and $u^2 = z$, where $u$ and $v$ are implicitly existentially
quantified. In turn, both equalities can be conjoined into a single
one by writing \mbox{$(uv-1)^2 + (u^2-z)^2 = 0$}, etc. We therefore
have:

\begin{proposition}
A graph $G$ is embeddable iff the polynomial $P_G$ has a real root.
\end{proposition}

It is easy to see that we can arrange for $P_G$ to have degree
four. Moreover, graphs with at most 30 vertices give rise to
polynomials in fewer than 1000 variables (the bulk of which are
required to ensure that all vectors are pairwise
distinct). Unfortunately, deciding whether such polynomials have real
roots is in general also well beyond the practical capabilities of
today's algorithms and computers. For an in-depth account of relevant
algorithms and results in this area, we refer the reader
to~\cite{ARAG}.

Finally, let us remark that graph embeddability can alternatively be
phrased in terms of \emph{isometric} (i.e., \emph{distance-preserving})
\emph{embeddability}: a graph is embeddable (in the sense of this paper)
iff its vertices can distinctly be placed in the upper half of
$\mathbb{R}^3$ so as to lie at distance 1 from the origin, and such
that adjacent vertices are precisely $\sqrt{2}$ units apart. More
information on isometric embeddings and related topics in topological
graph theory can be found in~\cite{LW09}.\vspace{1ex}

\noindent
\textbf{Homotopy Continuation.}  Homotopy continuation is a rigorous
yet fairly practical approach to solving systems of polynomial
equations over the reals. We experimented extensively with
Bertini~\cite{bertini}, a state-of-the-art software package for
numerical algebraic geometry. We briefly describe below some of the
algorithms underlying Bertini; for an authoritative and thorough
treatment we refer the reader to~\cite{Sommese05thenumerical}.

The general idea of homotopy continuation is the following: Given a
system of polynomial equations $\{f_i({\overline{x}}) = 0\}$ in variables
$\overline{x}$ that we are trying to solve (the `target' system), find
another system of equations $\{g_i(\overline{x}) = 0\}$ with known
solutions and of matching degrees (the `start' system).  Consider the
system of equations $\{h_i(\overline{x}, t) = t g_i(\overline{x}) + (1 -
t) f_i(\overline{x}) = 0\}$.  We know the solution of $\{h_i(\overline{x},
t) = 0\}$ at $t = 1$ and are seeking a solution at $t = 0$.  Assuming
that `nothing goes wrong' (which is a strong assumption,
see~\cite{Sommese05thenumerical} for details), we can numerically
track the paths of the roots of $\{h_i(\overline{x}, t)\}$ as $t$ goes
from $1$ to $0$.  This can be done, for example, by using Euler
prediction and Newton correction steps.

In order for this process to be successful, it is necessary to work
over complex rather than real numbers. When combined with several
enhancements, such as the \emph{gamma trick}, which makes use of a
random number generator in order to avoid bad start systems with high
probability, the method finds all the isolated roots of the target
system $\{f\}$, with a probability that can be made arbitrarily high.
If $\{f\}$ has only isolated roots and the number of these equals the
number of roots of $\{g\}$, then the method not only finds numerical
approximations to all the roots of $\{f\}$, but also in doing so,
demonstrates that there are no non-isolated roots.  To get a match in
the number of roots of $\{f\}$ and $\{g\}$, it can be helpful to
consider multi-degree structures~\cite{Sommese05thenumerical}. At the
conclusion of the continuation, it is then necessary to check which of
the roots thus obtained are indeed real.

Note that orthogonality constraints are homogeneous. Since Bertini
allows the definition of homogeneous groups of variables, it is
possible to work directly in the complex projective plane
$\mathbb{CP}^2$ and soundly dispense with all inequalities as well as
the equations requiring vectors to have unit length.

In order for a constraint system to have isolated solutions, it is
necessary to quotient out rotational symmetries, for example by fixing
three mutually orthogonal vectors. We must then have precisely as many
equations as unknowns.\footnote{Continuation also has techniques to
deal with over-constrained systems and non-isolated solutions, but the
options available in the current release of Bertini could not handle
these for the systems considered here.}

Any embedding found by Bertini should be considered a \emph{numerical}
existence proof, due to the use of floating-point numbers. Of course,
such solutions are however a very strong indication that a genuine
embedding does exist, which can then potentially be confirmed by more
expensive but exact methods. Moreover, under certain conditions, if
Bertini asserts that the solution set is empty then this should also
be considered a \emph{numerical} proof (of unembeddability), valid
with high probability.

In our experiments, Bertini was found to be relatively efficient on
small systems of equations. Unfortunately, its performance quickly
degraded with larger systems. For example, it was unable to solve (a
subset of) the constraints arising from the Conway-Kochen vector
system given in Fig.~\ref{fig_conway2} within a week's running 
time.\vspace{1ex}

\noindent
\textbf{Interval Arithmetic.} Interval arithmetic is a well-known
methodology for performing mathematically reliable numerical
computations. It is used by Pavi\v{c}i\'{c} \emph{et al}.\
in~\cite{pavicic,pavicic-2005-38} to tackle very similar embeddability
problems to the ones we consider here. Comprehensive treatments of the
subject can be found in a number of excellent texts such
as~\cite{MKC09}.

The basic idea underlying interval arithmetic is to replace precise
valuations of the variables with intervals in which a solution, if it
exists, must lie. Interval manipulations are performed by operations
on interval bounds that can even take into account rounding errors due to
floating-point arithmetic. 

A collection of intervals, one for each variable appearing in a system
of equations, is called a \emph{box}. Various techniques can be used
to progressively decrease the size of boxes, zeroing in on a
solution. When this process is exhausted, boxes are split into smaller
sub-boxes (a technique known as `bisection') and each case is analysed
separately.

Under the right circumstances, interval arithmetic can provide a
genuine, \emph{computer-aided} proof---and not merely a
\emph{numerical} proof---that a given graph is unembeddable: this
requires covering the entire space of potential solutions with a
finite number of (possibly very small) boxes, for each of which the
system of equations provably has no solution. Conversely, an interval
arithmetic solver can also produce \emph{computer-aided} proofs of
embeddability by proving that a root must lie in a given box. This can
be achieved through several means, the most well-known of which
employs an analogue of Newton's method for intervals. Of course,
interval arithmetic can also produce \emph{numerical} proofs of both
embeddability and unembeddability by narrowing in sufficiently closely
on particular regions of the state space.

Our implementation of an interval arithmetic solver made heavy use of
routines from the PROFIL library~\cite{profil}. Its performance varied
greatly, but it was able to produce proofs of both existence and
absence of embeddings for many of the graphs we experimented
with. Among others, it found a valid embedding for the graph
underlying the 31-vector Conway-Kochen system in a matter of seconds.

As explained in~\cite{Arends09alowerbound}, graphs that have a large
number of different 3-colourings will automatically possess a large
number of degenerate solutions (in which distinct vertices are mapped
to collinear vectors). Such solutions are especially difficult for an
interval analyser to handle, as they require a large number of correct
bisection decisions to be made and moreover produce a huge number of
boxes. Fortunately, as noted earlier, 3-colourable graphs are
\emph{ipso facto} 101-colourable, and are therefore not candidates for
producing KS vector systems in the first place. Since
101-colourability is in practice dramatically easier to establish than
unembeddability, this potential drawback is largely circumvented in
practice.\vspace{1ex}

\noindent
\textbf{Cubic Grids.}  Conway and Kochen's KS vector system of size 31
lies on a regular cubic grid, as shown in Fig.~\ref{fig_conway2}. That
grid can be viewed as consisting of all vectors with integer
coordinates lying on the surface of the cube $[-2,2]^3$, with
antipodal points identified.

We can, of course, consider grids of different granularities by
introducing a grid parameter $N$: the corresponding grid can be viewed
as the set of vectors with integer coordinates lying on the surface of
the cube $[-N,N]^3$, again with antipodal points identified.

One of the chief advantages of cubic grids is that all orthogonality
relationships are inferable by straightforward inspection; thus grid
embeddability provides a genuine mathematical proof of
embeddability. Moreover, we have so far not encountered any graph
which we believed to be embeddable (through the use of homotopy
continuation or interval arithmetic) yet which was not found 
to be embeddable on some cubic grid. This leads us to formulate the
following:

\begin{conjecture}
\label{conj}
Every embeddable graph can be embedded on some cubic grid.
\end{conjecture}

Note that a graph is embeddable on some cubic grid iff it has an
embedding on the surface of the unit cube $[-1,1]^3$ using only
rational coordinates. Interestingly, an assertion analogous to
Conjecture~\ref{conj} concerning the hemisphere $\mathbb{H}^2$ turns
out to be false: as shown in~\cite{Meyer99finiteprecision}, the set of
rational vectors on $\mathbb{H}^2$ (which is dense) \emph{can} be
101-coloured. Thus no KS vector system can possibly be represented on
$\mathbb{H}^2$ using exclusively rational coordinates.

In our experiments, we found grid-solving to be consistently highly
efficient. For example, embedding the 31-vertex Conway-Kochen graph
took less than 10ms on the grid with parameter $N=2$, approximately
250ms on the $(N\mbox{$=$}8)$-grid, and 26s on the
$(N\mbox{$=$}12)$-grid. Naturally, all embeddability proofs carry
mathematical certainty; however the absence of a grid embedding does
not allow one to draw any conclusion regarding (proper) embeddability.\\

Let us conclude this section by noting that embeddability clearly
remains a highly challenging problem at present, and seemingly one of
the hardest obstacles to obtaining tighter results than we have
been able to achieve to date.

\section{Lower Bounds}
\label{chap_lower_bound}

A natural strategy for finding small KS vector systems is first to
search for small graphs that are not 101-colourable. Such graphs
should moreover be square-free (otherwise they cannot be embeddable)
and connected (otherwise a smaller instance would be available).

Our initial approach was to generate these graphs at random, subject
to various parameters, and check whether they are embeddable. Several
hundred millions of connected square-free graphs were generated,
yielding thousands that were not 101-colourable. Unfortunately, for
most graphs with 30 vertices or less, we were simply unable to
determine embeddability; and the ones for which we did succeed were
all found to be unembeddable. Interestingly, our random graph
generator produced several isomorphic copies of the 31-vertex
Conway-Kochen specimen.

We then turned to sub-systems of the various cubic grids, which are
embeddable by construction. We were able to exhaustively search the
grids with parameters $N =2$ and $N=4$; all vector systems of size 30
or less were found to be 101-colourable, whereas the only systems of
size 31 that were not 101-colourable were all isomorphic to the
31-vector Conway-Kochen system. We also randomly sampled extensively
from sub-systems of the grids with parameters $N=6$, $N=8$, and
$N=12$.\footnote{For odd values of $N$, it turns out that the smallest
grid which is not itself 101-colourable---and therefore a candidate
for hosting KS vector systems---is the one with parameter $N=15$.} Again, no
smaller system was found, and all KS systems of size 31 were found to
be isomorphic to the Conway-Kochen system.\vspace{1ex}

\noindent
\textbf{Enumerating Connected Square-Free Graphs.}  At the time of
writing, the On-Line Encyclopedia of Integer Sequences~\cite{OEIS}
lists the numbers of non-isomorphic connected square-free graphs with
up to and including 17 vertices: there are 19,297,850,417 in total,
and 17,992,683,043 on 17 vertices alone. We re-enumerated all these
graphs and checked each one for 101-col\-our\-abil\-i\-ty, a task
which required solving more than 19 billion instances of an
NP-complete problem.

In~\cite{colbourn-1979a}, Colbourn and Read propose an `orderly'
procedure for graph enumeration. The key idea is to generate the
adjacency matrices of graphs in unique canonical forms. More
precisely, given an adjacency matrix, consider the bit-string obtained
by concatenating the entries strictly above the diagonal, column by
column (from top to bottom), left to right. The \defemph{canonical}
representation of a given graph $G$ is the unique adjacency matrix of
$G$ with the greatest bit-string value in lexicographic order.

As pointed out in~\cite{colbourn-1979a}, a crucial property of this
particular notion of canonicity is the following: if $\mathbf{M}$ is
the canonical adjacency matrix of a graph $G$ on $n$ vertices, then
the $(n-1) \times (n-1)$ submatrix of $\mathbf{M}$ obtained by
deleting the last column and the last row of $\mathbf{M}$ is also the
canonical adjacency matrix of some subgraph of $G$ on $n-1$
vertices. In the terminology of~\cite{Read78} (see
also~\cite{McKay98}), this enables the design of an \emph{effective}
graph enumeration algorithm: Generate the adjacency matrices of graphs
on a fixed number of vertices by a depth-first search process which
starts from the trivial $1 \times 1$ matrix and successively augments
the matrix by adding a single column and row to it until the target
number of vertices has been reached. In so doing, whenever a
non-canonical matrix is encountered, immediately discard it and
backtrack. This procedure guarantees that every canonical matrix will
appear exactly once at some point in the search. Moreover, the number
of non-canonical matrices that are produced (and immediately
discarded) in the process is kept relatively low.

A second key advantage of the Colbourn-Read notion of canonicity is
that one may use it to enumerate all non-isomorphic graphs with some
hereditary property (i.e., any property of a graph which automatically
holds for all its induced subgraphs). Note however that while
square-freeness is clearly hereditary, connectedness is
not. Nevertheless, the following result shows that the Colbourn-Read
algorithm is still suitable for our purposes:

\begin{proposition}
\label{connectednessresult}
Suppose that $\mathbf{M}$ is the canonical adjacency matrix of a
connected graph $G$. Then the submatrix of $\mathbf{M}$ obtained by
deleting the last column and the last row of $\mathbf{M}$ is also the
(canonical) adjacency matrix of some connected subgraph of $G$.
\end{proposition}

To see this, consider a graph $H$ over $n$ vertices having at least
two components, with canonical matrix $\mathbf{M}$. We claim that all
the vertices of one of its components will appear first in
$\mathbf{M}$. For clarity, let us list the vertices of $H$ as
$1,2,\ldots, n$ in the order in which they appear in $\mathbf{M}$. Let
$A$ be the component of $H$ which contains vertex $1$, and let $i$ be
the smallest vertex that does not belong to $A$. Suppose, for a
contradiction, that vertex $j < i$ is connected to some vertex greater
than $i$, which we will call $k$. Note that all $i-1$ entries of the
$i^\mathrm{th}$ column above the diagonal are 0. However, the
$j^\mathrm{th}$ entry of the $k^\mathrm{th}$ column is 1 (since $j$ is
connected to $k$) and therefore by swapping vertices $i$ and $k$
(which would not affect any of the entries above the diagonal to the
left of $i$) one would obtain an adjacency matrix for (a graph
isomorphic to) $H$ with strictly greater bit-string value than
$\mathbf{M}$, contradicting the canonicity of $\mathbf{M}$.

We now show that the last vertex (call it $n$) in the canonical
representation $\mathbf{M}$ of a connected graph $G$ is not a cut
vertex. Indeed, if removing $n$ disconnects $G$, then $G - \{n\}$ has
at least two components, and moreover there is an edge in $G$ from $n$
to each of the components of $G - \{n\}$. Recall however that the
submatrix of $\mathbf{M}$ obtained by deleting the last column and row
of $\mathbf{M}$ is a canonical representation of $G -
\{n\}$. Therefore all the vertices of some component $A$ of $G -
\{n\}$ appear first in $\mathbf{M}$. Moreover the first vertex $i$ not
belonging to $A$ has all entries above the diagonal set to 0. Lastly,
since $n$ is connected in $G$ to some vertex $j < i$, swapping
vertices $i$ and $n$ would yield an adjacency matrix for $G$ with
strictly greater bit-string value than $\mathbf{M}$, contradicting the
canonicity of $\mathbf{M}$. This concludes the proof of
Prop.~\ref{connectednessresult}.\qed

A major attraction of orderly graph enumeration algorithms is that
``\emph{expensive isomorphism tests are replaced by relatively
inexpensive verifications of canonicity}''~\cite{colbourn-1979a}---see
also~\cite{Read78,Read81,McKay98}. Canonicity checking is indeed a
pivotal component of the Colbourn-Read algorithm, yet somewhat
surprisingly its precise complexity appears to have remained open
since orderly algorithms were first introduced 30 years
ago.\footnote{Note that the problem of \emph{canonisation}---i.e.,
given a graph, construct its canonical adjacency matrix---is
well-known to be both NP-hard and co-NP-hard~\cite{BL83}. Yet it is
conceivable that merely \emph{verifying} canonicity could be
substantially easier.} We provide some partial answers to this
question below.

The first observation is that canonicity checking is clearly in co-NP:
if a matrix is not canonical, then one needs only exhibit another one
that is higher in lexicographic order together with an isomorphism
between the two. We now establish a hardness result:

\begin{theorem}
\label{hardnesscanonicity}
If the problem of canonicity checking were in NP, then NP =\linebreak 
co-NP.
\end{theorem}

We show that an NP algorithm for canonicity checking would entail the
existence of an NP algorithm for proving that a graph has no clique of
size a given integer $k$. Since the latter is well-known to be
co-NP-complete, the conclusion that NP = co-NP would follow.

Note that the canonical adjacency matrix of a graph having a clique of
size $k$ will necessarily contain 1s in the upper-left triangle
covering vertices up to $k$, for otherwise an adjacency matrix with a
higher lexicographic order could immediately be obtained by
re-labelling the vertices of the clique with the integers 1 to
$k$. Conversely, any adjacency matrix with an upper-left triangle
covering vertices up to $k$ consisting entirely of 1s necessarily
represents a graph having a clique of size $k$.

Let $G$ be a graph that has no clique of size $k$. Guess the canonical
adjacency matrix for $G$, guess and verify the adjacency mapping
showing that the matrix does indeed represent $G$, and verify that the
matrix is indeed canonical using the putative NP algorithm for
canonicity checking. By the above observation, the upper-left triangle
of this matrix covering vertices up to $k$ cannot consist entirely of
1s, thereby proving that all cliques in $G$ must have size strictly
less than $k$. \qed

Thm.~\ref{hardnesscanonicity} strongly suggests that canonicity
checking is unlikely to be in NP, and much less in P\@.

Note however that this hardness result is not obviously applicable to
square-free graphs, since in particular the latter have no cliques of
size greater than 3. We are nonetheless able to establish the
following weaker statement:

\begin{theorem}
\label{hardnesscanonicitysquare-free}
If the problem of canonicity checking for adjacency matrices of
square-free graphs were in NP, then the graph isomorphism problem (for
arbitrary graphs) would be in co-NP.
\end{theorem}

To prove Thm.~\ref{hardnesscanonicitysquare-free}, one first notices
that the graph isomorphism problem for square-free graphs is
polynomial-time equivalent to the graph isomorphism problem for
arbitrary graphs: given arbitrary graphs $G$ and $H$, transform both
into square-free graphs $G'$ and $H'$ by subdividing every edge
exactly once. Then $G$ and $H$ are isomorphic iff $G'$ and $H'$ are
isomorphic.

Suppose now that there were an NP algorithm for checking canonicity of
adjacency matrices of square-free graphs. Given two non-isomorphic
graphs $G$ and $H$, one could produce a certificate of non-isomorphism
by exhibiting the canonical adjacency matrices of $G'$ and $H'$ (as
defined above), together with the mappings showing that these matrices
are indeed adjacency representations of $G'$ and $H'$
respectively. Finally, validate that these matrices are indeed
canonical using the putative NP algorithm, and verify by inspection
that the two matrices differ. This proves that $G$ and $H$ are not
isomorphic and concludes the proof of
Thm.~\ref{hardnesscanonicitysquare-free}. \qed

We conjecture that both canonicity checking and canonicity checking
for square-free graphs are in fact co-NP-complete. Note that the
related but considerably more general problem of determining whether a
string is lexicographically maximum amongst all strings in its orbit
under a given arbitrary permutation group of the letters' positions is
known to be co-NP-complete~\cite{Jun01}. However it does not seem
possible to import that paper's techniques to our setting since the
corresponding permutation group acting on the set of entries of an
adjacency matrix is induced by the full permutation group on the
labellings of the vertices of the graph and as such is highly
constrained.

\begin{figure}
  \begin{center}
    \includegraphics[bb=104.375 5.5625 458.813 326.75,scale=.49]
{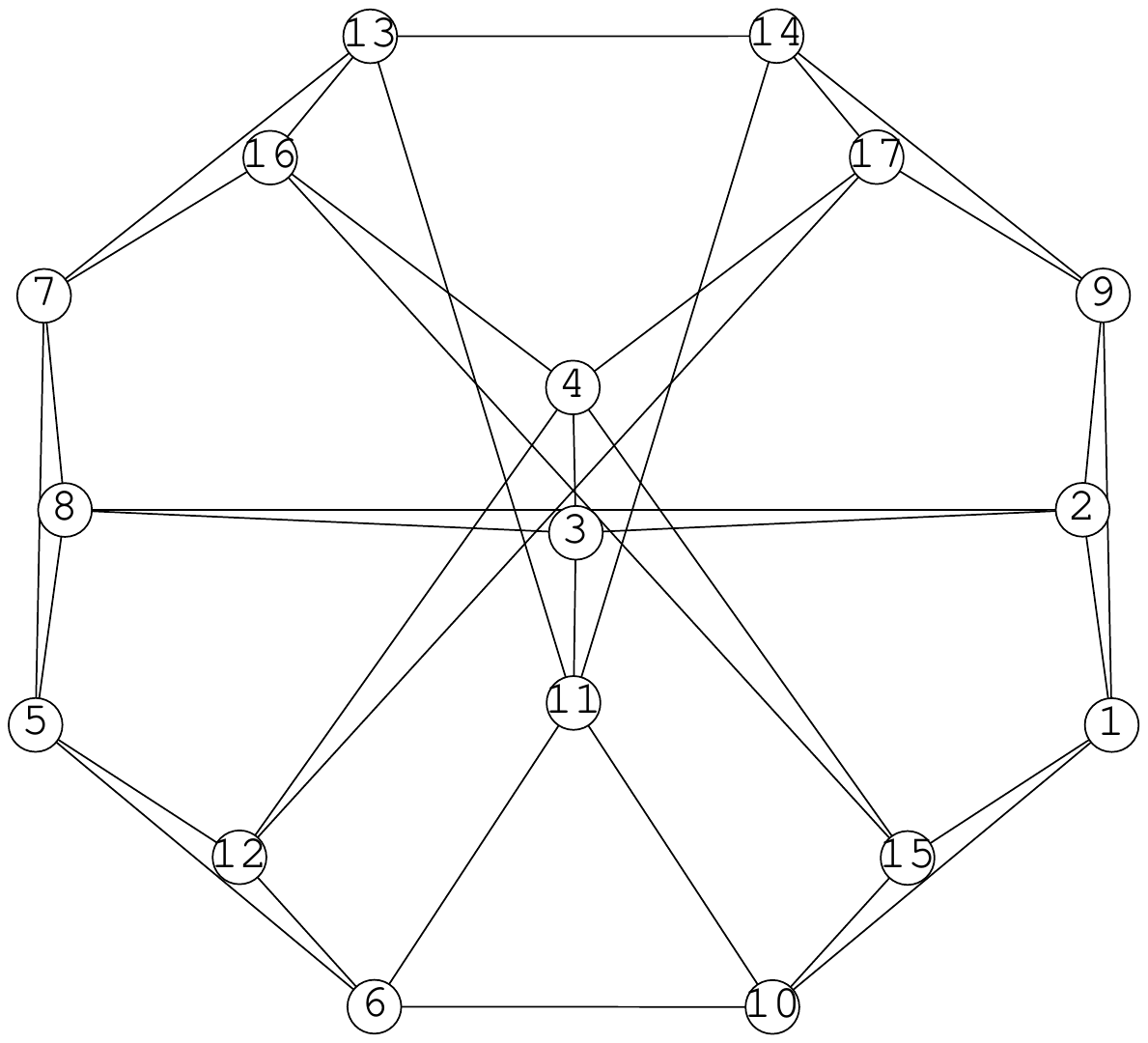}
   \includegraphics[bb = 91.5625 3.1875 436.313 306.688, scale=.49]
{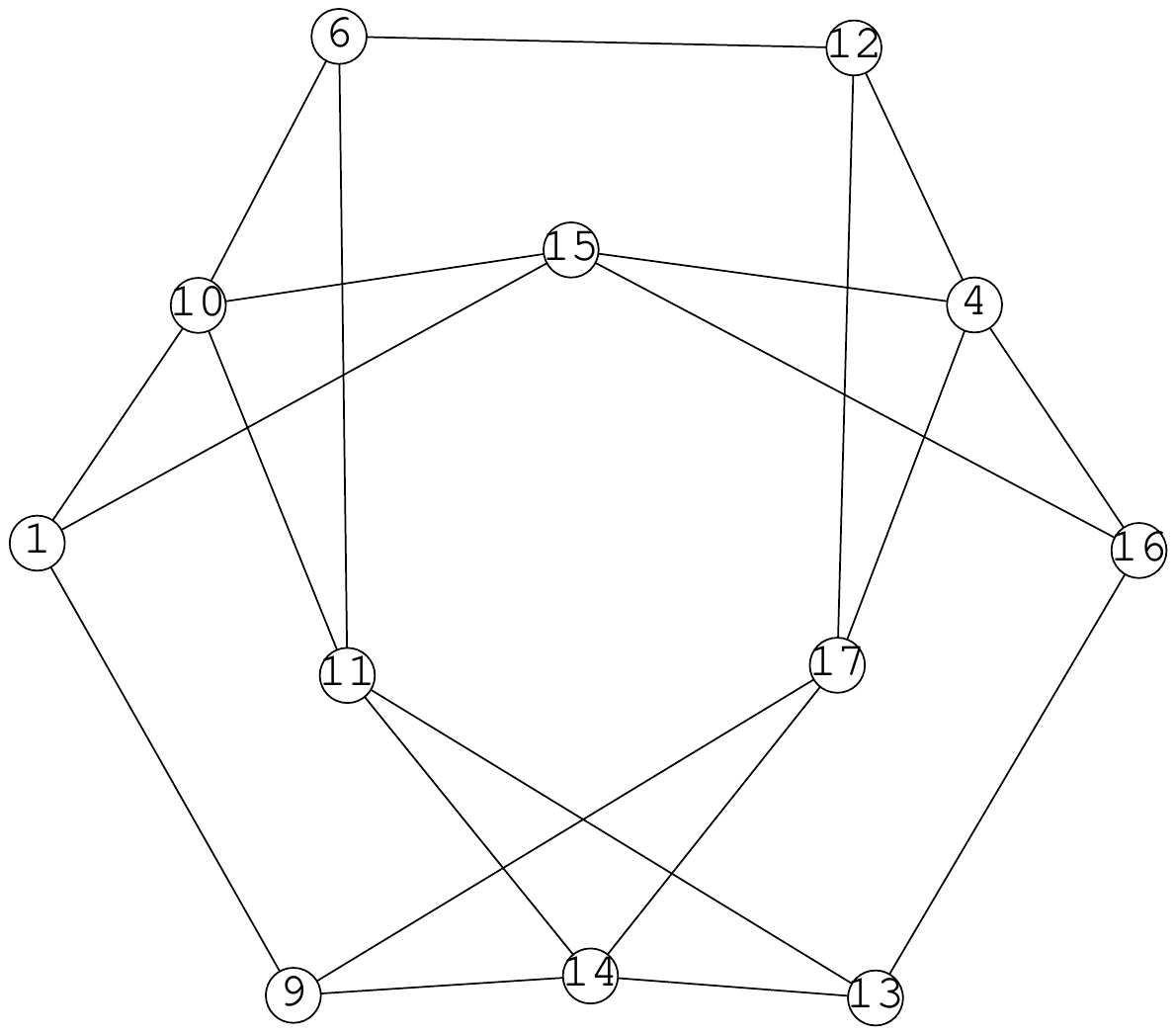}
  \end{center}
  \caption{\label{fig_uncolourable_17} 
The unique square-free graph on 17 vertices that is not
101-colourable (left), together with its smallest unembeddable subgraph.}
\end{figure}

In practice, notwithstanding Thms.~\ref{hardnesscanonicity} and
\ref{hardnesscanonicitysquare-free}, we have found that canonicity
checking could be made extremely efficient. We implemented a fairly
simple backtracking algorithm which enabled us to check canonicity of
the vast majority of matrices with 17 vertices or fewer within
microseconds. This led us to the following \emph{computer-aided}
result:

\begin{theorem}
Every square-free graph with at most 16 vertices is
101-colourable. Moreover, there is a unique graph with 17 vertices
that is not 101-colourable (shown on the left-hand side of
Fig.~\ref{fig_uncolourable_17}).
\label{thm_colourable}
\end{theorem}

Our next task was naturally to determine whether this 17-vertex
candidate is actually embeddable. Unfortunately, the answer is
no. Using Bertini, we identified a 12-vertex subgraph (shown on the
right-hand side of Fig.~\ref{fig_uncolourable_17}) whose associated
embedding polynomial was shown to have precisely 12 distinct complex
roots, none of which are purely real. This produces a \emph{numerical}
proof of the fact that our 17-vertex candidate cannot be
embedded. Finally, we were able to upgrade this to a
\emph{computer-aided} proof using our interval arithmetic solver,
yielding the following lower bound:

\begin{theorem}
  A Kochen-Specker vector system must contain at least 18 vectors.
  \label{lowerbound18}
\end{theorem}

Unfortunately, Thm.~\ref{lowerbound18} still leaves an astronomical
gap to bridge before proving that Conway and Kochen's KS vector system
of size 31 is the smallest possible (if indeed that is the
case). Extrapolating from known data, the graph below suggests that
there are some $10^{32}$ connected square-free graphs on 30 vertices
or less, well out of the brute-force reach of current technology.

\begin{center}
\includegraphics[scale=0.6]{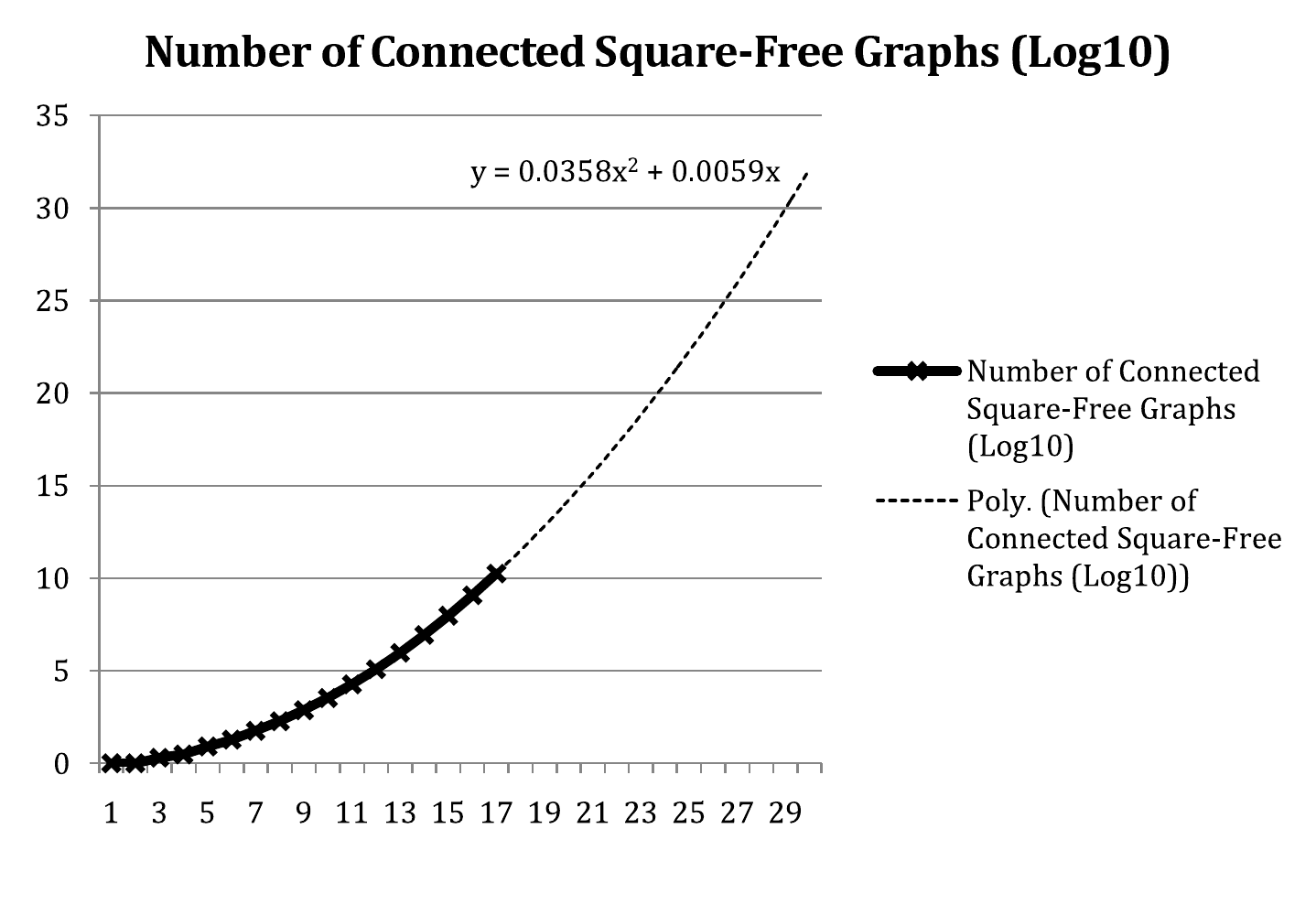}
\end{center}

\section{Conclusion}

We have proposed the problem of finding small Kochen-Specker vector
systems---or proving that none exist of size less than 31---as a
difficult and worthwhile algorithmic challenge. Higher-dimensional
generalisations of the problem have also been considered by others,
notably Pavi\v{c}i\'{c} \emph{et al.}~\cite{pavicic-2005-38}.

The results we have obtained (greater details of which are available
in~\cite{Arends09alowerbound}) can largely be summarised by listing a
number of properties that any minimal KS vector system must
enjoy. Such a system:
\begin{itemize}
\item has at most 31 vectors (Conway and Kochen's KS vector system);
\item contains at least 18 vectors (Thm.~\ref{lowerbound18});
\item has associated graph that is square-free (Sec.~\ref{secKS});
\item has associated graph that is not 101-colourable and not
3-colourable (Sec.~\ref{secKS});
\item has associated graph that is 4-colourable 
(Prop.~\ref{prop4colourable});
\item has associated graph with minimum degree
  3~\cite{Arends09alowerbound};
\item has associated graph in which each vertex belongs to a
      triangle~\cite{Arends09alowerbound};
\item is not a subsystem of the cubic grid with grid parameter $N =
      4$, unless it is the Conway-Kochen 31-vector system itself
      (Sec.~\ref{chap_embeddability}).
\end{itemize}

In our view, two central challenges are to (i)~devise more efficient
algorithms for determining graph embeddability (in which respect
Conjecture~\ref{conj} could play a key role), and (ii)~find efficient
means to drastically cut down the number of candidate graphs that must
be examined.\vspace{1ex}

\noindent
\textbf{Acknowledgements.} We thank Nick Trefethen for introducing the
first two authors to the third, Jean-Pierre Merlet for sharing his
experience on interval arithmetic, and Don Knuth for drawing our
attention to~\cite{colbourn-1979a}. Following publication of the short
version of this paper~\cite{AOW11a}, Ad\'an Cabello informed us that
the unique square-free uncolourable graph on 17 vertices depicted in
Fig.~\ref{fig_uncolourable_17} had already been discovered by John
Smolin several years earlier. To the best of our knowledge, this graph
was never published, nor were the methods used to obtain it
disclosed. Cabello further claims in~\cite{Cab06} to have proof that
this graph is unembeddable, although the reference he gives in
\emph{op.~cit.} is again unpublished.

The second author was supported by EPSRC and the third author was
supported by NSF grant DMS-0712910.

\bibliography{wg11b}

\end{document}